\documentclass[twocolumn,aps,prl,showpacs]{revtex4}

\usepackage{graphicx,color,hyperref}

\usepackage{amsmath}

\usepackage{amssymb}

\definecolor{orange}{rgb}{1,0.5,0}
\definecolor{grey}{rgb}{.6,.6,.6}

\begin{document}

\title{Strong Anderson localization in cold atom quantum quenches}

\author{T.~Micklitz$^1$, C.~A.~M\"uller$^2$, and A.~Altland$^3$}

\affiliation{$^1$Centro Brasileiro de Pesquisas F\'isicas, Rua Xavier Sigaud 150, 22290-180, Rio de Janeiro, Brazil \\
$^2$Fachbereich Physik, Universit\"at Konstanz, 78457 Konstanz, Germany\\
$^3$Institut f\"ur Theoretische Physik, Universit\"at zu K\"oln, Z\"ulpicher Str. 77, 50937 Cologne, Germany}

\date{\today}


\begin{abstract}
Signatures of  Anderson localization  in the momentum distribution of a
cold atom cloud after a quantum quench are studied. We consider a quasi one-dimensional 
cloud  initially prepared in a well defined momentum state, and
expanding for some time in a disorder speckle potential. Quantum interference 
generates a peak in the forward scattering
amplitude which, unlike the common weak localization backscattering peak, is a signature of 
\emph{strong} Anderson localization.
We present a non-perturbative, and fully time resolved description of the
phenomenon, covering the entire diffusion--to--localization crossover. Our
results should be observable by present day experiments. 
\end{abstract}

\maketitle

After several decades of research there remains a 
stark imbalance between a huge body of theory  
and scarcely any
controlled experimental observation 
of Anderson localization in generic disordered systems~\cite{footnote}. 
Here `controlled' means
the option to link experimental signatures directly to underlying quantum interference processes 
via a tuneable parameter. 
Ultra-cold atoms are likely our best bet 
to improve upon this situation, and experiments based on a quench protocol appear to be 
particularly promising. Within the quench paradigm, `time' plays the role of a control parameter, 
and Anderson localization would be monitored in the slow genesis of an observable strongly affected by 
disorder generated quantum interference. A specific proposal going in this direction has recently been 
made~\cite{Cherroret2012,Karpiuk2012} and realized~\cite{Josse,Labeyrie2012}.
These experiments expose a cloud of cold atoms with initially well defined momentum,
$\bold{k}_{\rm i}$, to a laser speckle disorder potential.
Suspended against gravity by magnetic levitation, the cloud propagates in the
disorder potential for some time, $t$, after which all potentials are turned
off and the atomic momentum distribution $\rho(\bold{k}_{\rm f},t)$ is
determined by time of flight measurement.

For this setup, theory~\cite{Karpiuk2012} predicts the presence of a forward 
scattering peak ($\bold{k}_{\rm f} \simeq
\bold{k}_{\rm i}$), besides the familiar weak localization backscattering 
($\bold{k}_{\rm f} \simeq -\bold{k}_{\rm i}$)
peak often observed in such types of experiment.
The remarkable
difference between the two structures is that the forward peak is a manifestation of
\emph{strong} Anderson localization, i.e. a non-perturbative
accumulation of quantum coherence processes.
First indications to  the emergence of a forward peak have been extracted  from an
insightful combination of perturbation and scaling theory in 
Ref.~\cite{Karpiuk2012}. However, the full profile of the signal, its height, width, and temporal
development, can only be addressed in terms of the non-perturbative methods tailored to the
description of strong localization phenomena~\cite{efetovbook,efetovreview}. 
At any rate, an observation of the peak formation --- which arguably is in reach of 
present experimentation --- and its successful comparison to
time-resolved analytic results would provide us with an exceptionally strong
testbed for our understanding of strong localization phenomena.
 
In this paper, we present a fully analytic theory of the forward scattering
peak in the quantum quench protocol. Particular attention is payed to the genesis of
the peak at the time scale characteristic for the buildup of
strong localization phenomena, 
an analysis made possible thanks to recent progress~\cite{localCorrelations}.
Our theory also describes the structure of the fully developed 
peak in momentum space, i.e. its height relative to the isotropic background,
its width, and its dependence on the experimentally unavoidable initial momentum spread. 

Below we consider a situation where time reversal ($\mathrm{T}$) invariance is
broken by a weak synthetic gauge field. 
In this case, the diffusion modes relevant to the formation of the backscattering peak 
(`Cooperon modes' in the parlor of the field) are frozen out, and the signatures of localization 
reside entirely in the forward peak. While the absence of Cooperon modes technically simplifies 
our analysis,  strong localization is only weakly affected --- e.g., the localization length doubles 
compared to the $\mathrm{T}$ invariant case --- and the  
features of the forward peak are not expected to be modified in  
essential ways.


{\it Model and effective theory:---}We consider a cloud of non-interacting atoms 
confined to a quasi one-dimensional geometry of extensions $L_{y,z}\ll L_x$, and  described by the
Hamiltonian $\hat{H}_0=(\hat{\bold{p}}-\mathbf{a})^2/ 2m$, where $\mathbf{a}$
is a weak $\mathrm{T}$--breaking synthetic gauge
field~\cite{SynGauge,WeakField}.  Assume the cloud to be initially prepared in a
momentum eigenstate $|\bold{k}_{\rm i} \rangle$; 
residual effects due to
momentum spread will be discussed below. At time $t=0$ we switch on a random
potential, $\hat H_0 \to \hat{H}\equiv \hat H_0 + V(\hat{\bold{x}})$. Assuming
$V$ to represent a short range laser speckle, we model it as  Gaussian white
noise, $\overline{V(\bold{x}) V(\bold{x}')} = {1\over
2\pi\nu\tau}\delta(\bold{x}-\bold{x}')$, where $\nu$ is the density of states
per volume, $\tau$ is the elastic scattering time, and $\hbar$ has been set to
unity.

The observable of interest is the time dependent configuration averaged fidelity
\begin{align}
\label{overlap}
{\cal C}_{\bold{k}_{\rm i} \bold{k}_{\rm f}}(t)
&=\overline{|\langle \bold{k}_{\rm f} |e^{-i\hat{H}t}|\bold{k}_{\rm i}\rangle|^2},
\end{align} 
between initial and final state, which is directly measurable by time of 
flight absorption measurement. At times larger than the mean free time, 
the response function ${\cal C}$ splits into a sum of two pieces, 
 \begin{align}  \label{Ckisopeak}
{\cal C}_{\bold{k}_{\rm i} \bold{k}_{\rm f}}
&\stackrel{t>\tau}{=}
{\cal N}  f_{\bold{k}_{\rm i}}f_{\bold{k}_{\rm f}}
\left[ {\cal C}_0
+  {\cal C}_1(\bold q)
  \right],
\end{align}
where $\mathcal{C}_0$ and $\mathcal{C}_1$ are the isotropic background and the
momentum dependent contribution to the correlation function, resp., where
$\bold{q}= \bold k_{\rm i} - \bold k_{\rm f}$ is the momentum difference, and
the weight function $f_{\bold{k}} =[1+ (2\tau ({\bold{k}^2\over
2m}-E))^2]^{-1}$ restricts initial and final momenta to a disorder-broadened
shell of energy, $E$,  and ${\cal N}$ guarantees the normalization
$\sum_{\bold{k}'} {\cal C}_{\bold{k}
\bold{k}'}(t)=1$.

Classical kinetic theory predicts that at time scales larger than the elastic
scattering time, $\tau$, the  momentum distribution will diffusively relax to
an isotropic configuration with $\mathcal{C}_1=0$ \cite{Plisson2013}.  Quantum
coherence introduces the Heisenberg time $t_H=1/\Delta_\xi$, where
$\Delta_\xi$ is the single particle level spacing of a `localization volume'.
In our quasi one-dimensional setup,  $\Delta_\xi  = (2\pi
\nu S\xi)^{-1}$, where $S=L_yL_z$ is the cross section of the system, and $\xi\gg
L_{y,z}$ the longitudinal localization length due to disorder. The latter can
be implicitly defined by the equality $\Delta_\xi=D/\xi^2$ 
to the inverse of the diffusion time through a localization volume as
$\xi=2\pi\nu SD$, where $D$ is the three-dimensional diffusion
constant on the energy shell.  We assume the system to be strongly localizing in the 
sense $\xi\ll L_x$, at negligibly weak finite size corrections in $\xi/L_x$. 
In the following we will construct a microscopic theory of the
appearance of this scale and its influence on the evolution of the forward
peak.

We start out by Fourier transforming the correlation function to frequency
space, ${\cal C}(t) = \int{d\eta \over 2\pi} {\cal C}(\eta) e^{-2it\eta_+}$,
where $\eta^+ = \eta+i0$. The function $\mathcal{C}(\eta)$ then assumes the standard
response form of a product of two single particle Green functions, 
which is to be averaged over disorder. Expressions of this
type are tailored to an analysis in terms of the supersymmetric
nonlinear $\sigma$-model, and a straightforward application of the formalism of 
Ref.~\cite{efetovbook} yields our momentum correlation functions as 
\begin{align}
\label{0ciso}
{\cal C}_0(\eta)
&=
\langle \,\mathrm{tr}\left(
 \mathcal{P}_{++}\, Q(0)\,\mathcal{P}_{--}\, Q(0) 
  \right)\rangle_{S_0}, 
\\
\label{0cfs}
{\cal C}_1(\bold q,\eta) 
&=\delta_{\mathbf{0},\bold q^{\perp}}
 \langle \mathrm{tr}\left(
 \mathcal{P}_{-+} \,Q(q)\, \mathcal{P}_{+-}\,
 Q(-q) \right)
\rangle_{S_0}.
\end{align}
Here $Q=\{Q^{\alpha\alpha'}_{ss'}\}$ is a $4\times 4$ supermatrix, 
that obeys the nonlinear constraint $Q^2=\openone$. Its diagonal $2\times 2$ 
upper- left and lower-right matrix blocks $Q^{\mathrm{bb}}$ and $Q^\mathrm{ff}$, resp., 
contain complex numbers, while the off-diagonal blocks $Q^\mathrm{bf,fb}$ contain Grassmann variables. 
The subscript indices $Q_{s,s'}$, $s,s'=\pm$ discriminate between `retarded' and
 `advanced' components of the matrix field, and the matrices $\mathcal{P}_{ss'}$ project on
 the $s,s'$ advanced/retarded block in the fermion-fermion sector. The
 correlation function \eqref{0cfs} preserves the initial transverse momentum,
 $\bold{q}^\perp=\mathbf{0}$, and is sensitive to the longitudinal momentum
 difference $q =k_{\rm i}^x - k_{\rm f}^x$.

\begin{figure}[tt]
\centering
\centerline{\includegraphics[width=9cm]{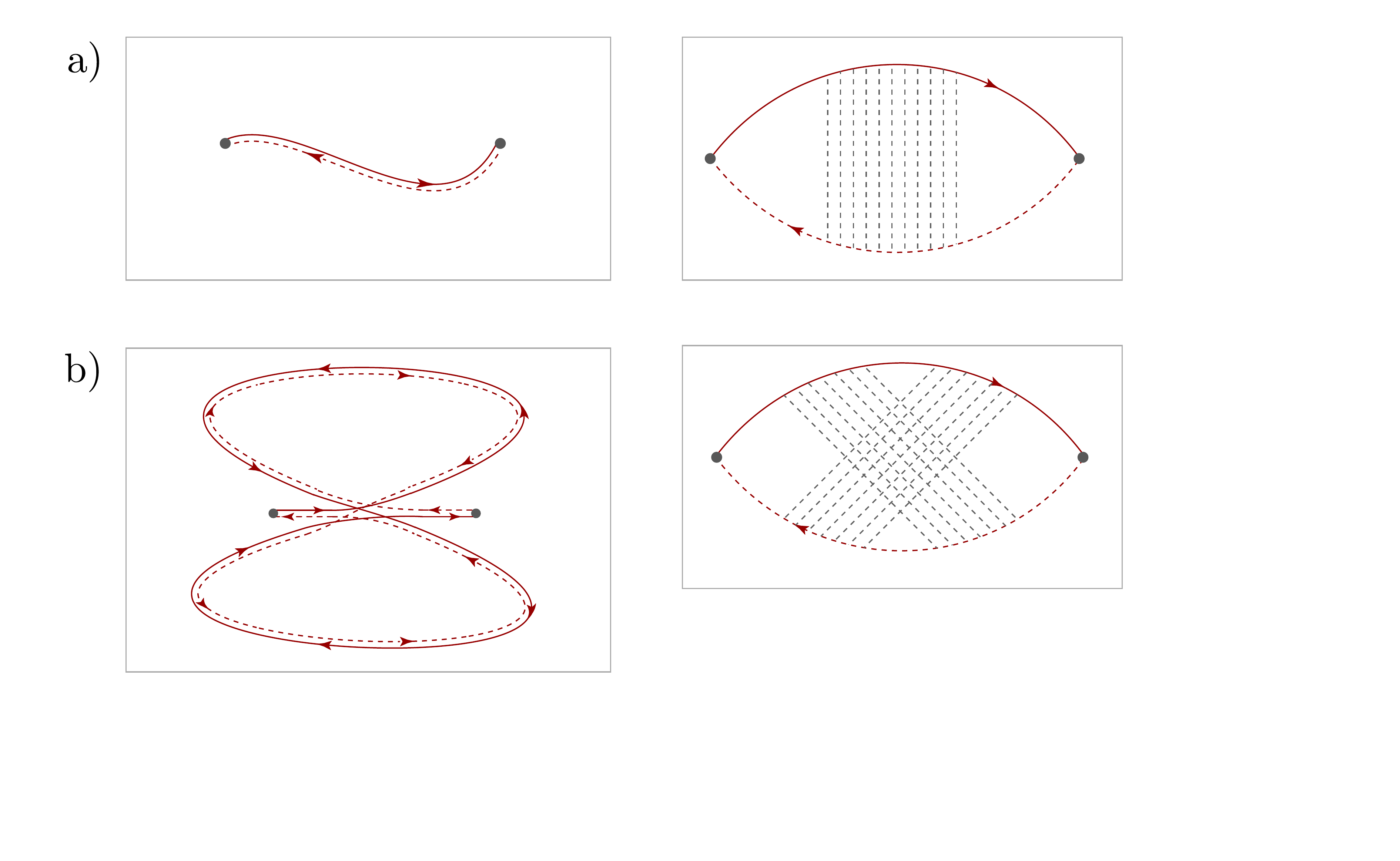}}
\vspace{-15pt}
\caption{Leading, a), and subleading, b), contributions to the correlation
function (1) in the perturbative limit. a) Classi\-cal diffusive propagation
of a particle-hole pair in real space (left panel), and the corresponding
impurity diagram in momentum space (right panel), which does not retain 
memory of the initial momentum and contributes to the isotropic background, 
Eq.~\eqref{0ciso}. b) Leading contribution to the forward correlation (4). A 
particle-hole pair amplitude splits up into two diffusive loops traversed in opposite
order (left panel). The ensuing quantum interference contribution to the
correlation function (right panel) does not rely on T-invariance and is
strongly peaked in forward direction, $q= 0$. }
\label{fig1}
\end{figure}

 Finally, the functional average in ~\eqref{0ciso} is defined as $\langle ...
 \rangle_{S_0}=\int {\cal D}Q e^{S_0[Q]}(\dots)$, where $S_0$ is the celebrated
 diffusive $\sigma$-model action
\begin{align}
\label{action}
S_0[Q]
&=  \pi\nu S \int dx\, 
{\rm str}\left(i\eta Q \Lambda
+
{D\over 4}(\partial_x Q)^2 \right),
\end{align}
where $\Lambda=\{s\, \delta_{ss'}\}$ is the identity matrix in boson-fermion space 
but breaks symmetry in advanced-retarded space.  
Throughout it will be convenient to think of the coordinate $x$ in
terms of some fictitious `time' and of $Q$ as a multi-dimensional quantum
particle. Eq.~\eqref{action} then acquires the status of a Feynman path
integral, with `kinetic' energy $\sim (\partial_xQ)^2$, and a `potential'
$\sim \eta Q \Lambda $. The latter  is invariant
under similarity transformations of $Q$ leaving the diagonal matrix $\Lambda$
invariant. As with a `radial' potential in quantum mechanics, this high degree
of symmetry will reduce the effective dimensionality of the problem and make
non-perturbative calculations possible.  Furthering upon this analogy, much of
our analysis  will rest upon a mapping of the path integral to a Schr\"odinger
equation, which can be  addressed in  analogy to the quantum
mechanics of centro-symmetric potentials. However, before turning to this
discussion, we address the short time, or `strong potential' limit, where
$\eta$ is large enough to confine the particle close to the origin and a
formulation in terms of `cartesian' coordinates is appropriate.


{\it Diffusive short time limit $\tau\ll t \ll t_H$:---}The dynamics  
on (real) time scales shorter than the Heisenberg time, or $\eta\gg
\Delta_\xi$, can be addressed in terms of perturbation theory around the
high-frequency saddle point $Q=\Lambda$ as $Q=T\Lambda T^{-1}\simeq \Lambda
(1-2W+2W^2 -\dots)$, where $T=\exp(W)$ and $W$ are fluctuation generators.
Individual terms in the ensuing perturbation series afford  an interpretation
in terms of the diffusive `ladder diagrams'~\cite{efetovbook} shown in Fig.~\ref{fig1}.
The leading $\mathcal{O}(W^2)$, or zero loop order [Fig.~\ref{fig1}, a)],
contributes to the isotropic part, ${\cal C}_0$, where it describes momentum
relaxation at time scales larger than the scattering time $\tau$. The first
contribution to the forward peak arises at $\mathcal{O}(W^4)$, or two-loop
order [Fig.~\ref{fig1} b)], as ${\cal C}_1(t)\propto\sqrt{t/t_H}$. This
increase is fast in comparison to the $2d$ scaling $\sim t/t_H$ obtained in
Ref.~\cite{Karpiuk2012} (for a time reversal invariant
setting \cite{fncrossdiff}), and reflects the relatively higher phase volume
accessible to diffusive fluctuations in low dimensions. The appearance of
$t_H$ as a reference scale indicates that localization is relevant to the phenomenon.

{\it Mapping to differential equations:---} For  larger time scales $t\sim t_H$, 
or `weak potential' $\eta \sim \Delta_\xi$, it is advantageous to switch from the path 
integral~\eqref{Ckisopeak} to an equivalent `Schr\"odinger equation'. It's  Hamilton operator,
\begin{align}\label{transferH0} 
 \hat{H}_0 
&=
-
 \left(  
 \partial_\lambda 
 {1-\lambda^2 \over \lambda_-^2} \partial_\lambda
 +
 \partial_{\lambda_1} 
 {\lambda_1^2-1 \over \lambda_-^2} \partial_{\lambda_1}
  \right)
- {i\over 2} \eta t_H \lambda_-,
\end{align}
couples to the two `radial' coordinates, $\lambda,\lambda_1$ of the 
quantum particle~\cite{efetovbook,efetovreview} through a kinetic energy  and a
potential term, resp., where $\lambda_-=\lambda_1-\lambda$. 
Central to our problem are the ground state wave
function, $\hat{H}_0 \Psi_0 = 0$, $\Psi_0=\Psi_0(\lambda,\lambda_1)$, with
boundary condition $\Psi_0(1,1)=1$, and a perturbed wave function obeying 
\begin{align}
\label{psi1}
\left( 2\hat{H}_0-i q\xi \right) \Psi_1^q
&= \lambda_- \Psi_0, 
\end{align}
where $q$ is the longitudinal momentum difference.
Applying the formalism of Refs.~\cite{efetovbook,efetovreview} the main observable of 
interest, Eq.~\eqref{0cfs}, can then be expressed as 
\begin{align}
\label{cfs}
{\cal C}_1(q,\eta)
&= 
\xi \int \frac{d\bar{\lambda}}{\lambda_-} 
\left[\Psi_1^q(\bar{\lambda})+\Psi_1^{-q}(\bar{\lambda})\right]
\Psi_0(\bar{\lambda}) , 
\end{align} 
where $\int d\bar{\lambda}=\int_1^\infty d\lambda_1\int_{-1}^1 d\lambda$. The
isotropic contribution $\mathcal{C}_0$ can be  described similarly, but its explicit representation
will not be needed here~\cite{details}.
The behavior of the `wave functions' entering~\eqref{cfs} 
depends on whether we are (i) in the
regime of short times $t\ll t_H$, (ii) the diffusion--to-localization crossover regime of
intermediate times $t\simeq t_H$, or (iii) at asymptotically long times $t\gg t_H$.  In (i), straightforward perturbation theory
in weak perturbations off the configuration $\lambda=\lambda_1=1$ pinned by
the strong potential $\eta \gg 1/t_H$ recovers the results summarized above~\cite{details}.
We now proceed to explore how the short time asymptotic connects to the perturbatively inaccessible crossover regime, (ii). 

\begin{figure}[tt]
\begin{center}
\includegraphics[width=8cm]{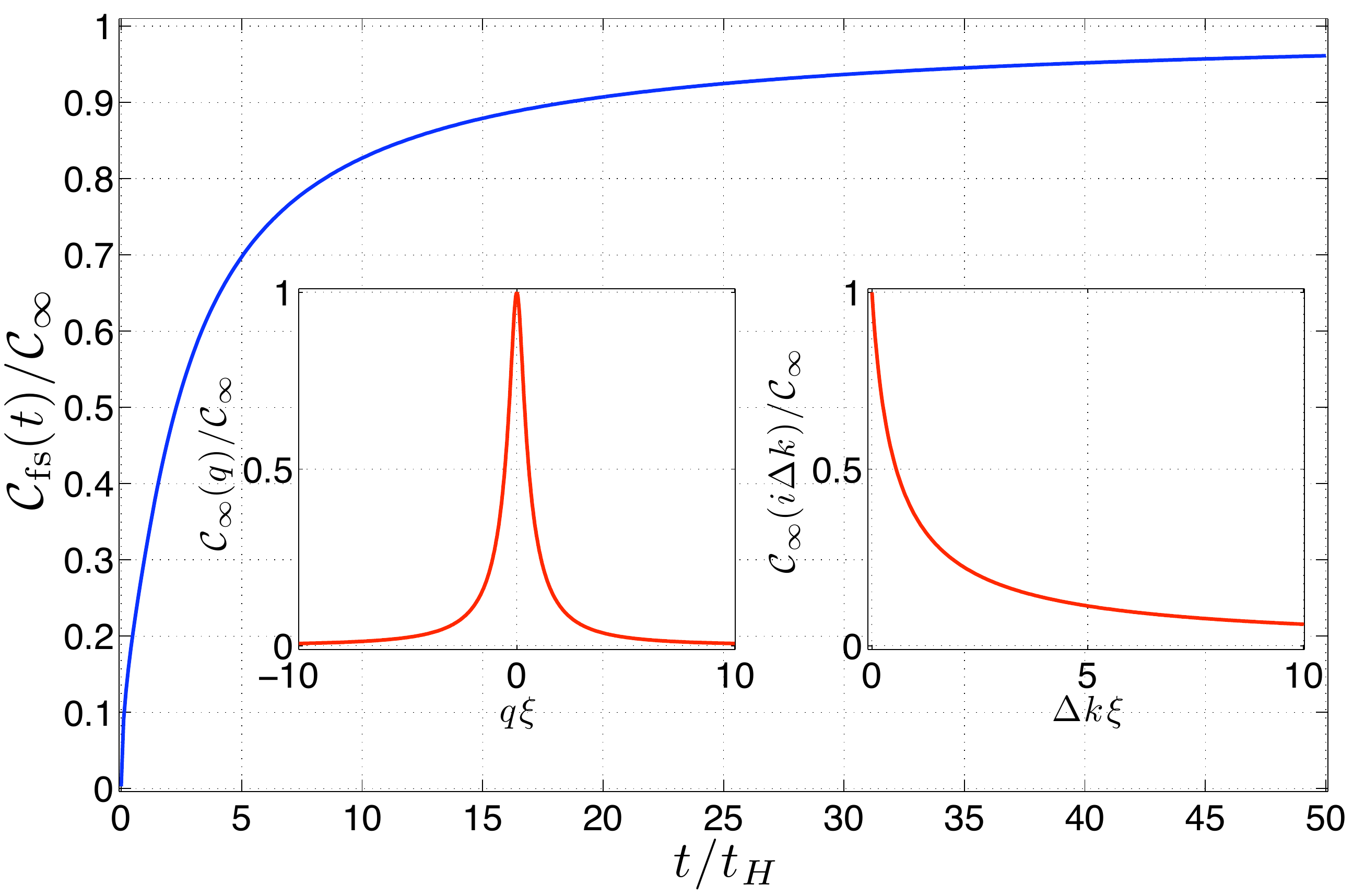}
\end{center}
\vspace{-15pt}
\caption{Forward-scattering peak contrast, Eq.~\eqref{fsp}, 
 as a function of time $t/t_H$. Insets: Saturation value 
 as a function of longitudinal momentum difference $q$ (left) and
 spread $\Delta k$ of the initial state (right).}
\label{fig2}
\end{figure}

{\it Diffusion-to-localization crossover:---} At intermediate times, the primary goal is a description of
the temporal buildup of the forward peak $\mathcal{C}_\mathrm{fs}(t)=\mathcal{C}_1(0,t)$. 
Building on recent progress
by Skvortsov and Ostrovsky (SO) \cite{localCorrelations} we will find that the
problem  possesses a surprisingly simple
solution. The key observation of SO was that upon introduction of elliptic
coordinates, $\lambda=\frac{1}{2}(r-r_1)$, $\lambda_1=\frac{1}{2}(r+r_1)$,
$r=\sqrt{z^2+\rho^2}$, and $r_1=\sqrt{(z-2)^2+\rho^2}$,  $\hat H_0$ assumes a
form similar to that of a  non-relativistic $3d$ Coulomb Hamiltonian,
\begin{align}
\hat{H}_0
&=-{r_1^2r\over 2}
\left[
\Delta_0
-
{2\kappa\over r}
\right] {1\over r_1}, 
\end{align}
where $\Delta_0\equiv \partial_z^2 + \frac{1}{\rho}\partial_\rho
\rho\partial_\rho$ is the  Laplace operator in cylinder coordinates $(\rho,\varphi,z)$ acting in
the space of azimuthally symmetric functions, and $\kappa\equiv -i\eta t_H/2$.
Following Ref.~\cite{localCorrelations}, we can derive the ground state wave function from the known
zero energy Green's function  of the $3d$ Coulomb-problem \cite{Meixner1933},
\begin{align}
\label{CoulombG}
\left[ 
\Delta_0 - {2\kappa\over r} 
\right] G_0(\bold{r},\bold{r}')
&=\delta(\bold{r}-\bold{r}'),
\end{align}
where $\bold{r}'=(0,\varphi,2)$ corresponds to the boundary point
$(\lambda,\lambda_1)=(1,1) \leftrightarrow(r,r_1)=(2,0)$. From this function,
we obtain a solution to the ground state problem as, $\Psi_0(r,r_1)=-4\pi
r_1G_0(\bold{r},\bold{r}')$, where the role of the
$\mathbf{r}'$-inhomogeneity in \eqref{CoulombG} is to implement the 
boundary condition $\Psi_0(0,2)=1$. A solution for the Green function in terms 
of Bessel functions has been obtained long ago~\cite{gf} as
\begin{align}
\label{gf}
G_0(\bold{r},\bold{r}')
&=
{ (\partial_u-\partial_v)\sqrt{u}K_1(2\sqrt{\kappa u})\sqrt{v}I_1(2\sqrt{\kappa v})
\over 2\pi |\bold{r}-\bold{r}'|},
\end{align}
where $u= r+r'+|\bold{r}-\bold{r}'|$ and $v= r+r'-|\bold{r}-\bold{r}'|$.
Thanks to the appearance of the Green function, the solution to our full
problem can now be formulated by standard methods of quantum mechanics. 
Specifically,  we observe that the
solution for the excited wave function \eqref{psi1} is obtained by convolution
of the Green function \eqref{gf} and the source term ${1\over
rr_1}\Psi_0(r,r_1)$ over the three dimensional volume element $d^3r=(r
r_1/2)dr dr_1d\varphi$. Substitution of this result into Eq.~\eqref{cfs} then yields 
\begin{align*}
{\cal C}_\mathrm{fs}(\eta)
&=
32\pi  \xi 
\langle
\bold{r}_0|
\hat{G}_0 {1\over \hat{r}} \hat{G}_0 {1\over \hat{ r }} \hat{G}_0
|\bold{r}_0
\rangle \stackrel{\eqref{CoulombG}}{=} 8\pi \xi \partial^2_\kappa G_0(\mathbf{r}_0,\mathbf{r}_0).
\end{align*}
To compute  the
$\kappa$-derivative, we regularize the Green function in \eqref{gf} as
$G_0(\mathbf{r}_0,\mathbf{r}_0)=\lim_{\mathbf{r}\to
\mathbf{r}_0}G_0(\mathbf{r},\mathbf{r}_0)$. A straightforward Taylor expansion
of  Bessel functions then leads to 
\begin{align}
\label{fspeta}
{\cal C}_\mathrm{fs}(\eta)
&=
\left. 8\xi \partial_\kappa \,
K_0(4\sqrt{\kappa})I_0(4\sqrt{\kappa})\right|_{\kappa = -i \eta \frac{t_H}{2}}. 
\end{align}
Upon Fourier transformation, the time dependence  of the
forward-scattering peak contrast is finally obtained as (cf. Fig.~\ref{fig2}) 
\begin{align}\label{fsp}
{{\cal C}_\mathrm{fs}(t)\over \mathcal{C}_\infty}
&=
\theta(t)
 I_0\left( {2t_H \over  t} \right)
e^{-2 t_H / t}, 
\end{align}
where $\mathcal{C}_\infty = \mathcal{C}_\text{fs}(\infty)$ is the long term asymptotic to be discussed momentarily. 
The limiting behavior of 
$I_0$~\cite{Gradsteyn} implies the long and short time expansions  
\begin{align}
{{\cal C}_{\rm fs}(t)
\over \mathcal{C}_\infty}
&=
\begin{cases}
{1\over\sqrt{2\pi}}\Big( \left(\frac{t}{2t_H}\right)^{1\over 2}
+{1\over 8}\left(\frac{t}{2t_H}\right)^{3/2}+\dots \Big), 	
&   t \ll t_H, 
\\
1-2\frac{t_H}{t}+3\left(\frac{t_H}{t}\right)^{2}+\dots, 
&   t\gg t_H, \nonumber
\end{cases}
\end{align}
generalizing the previously studied limit.

{\it Long time limit $t/t_H\gg 1$ and saturation value of contrast:---}For $\eta\ll t_H^{-1}$, 
fluctuations far away from the origin
$(\lambda,\lambda_1)=(1,1)$ become energetically affordable. In this regime,
dominant contributions to correlation functions come from the integration over
$\lambda_1\gg\lambda\sim 1$. To leading approximation the dependence of the
differential equations on $\lambda$ may  be
ignored~\cite{efetovbook,efetovreview}, and the solution for the dependence on
the `non-compact' variable $\lambda_1$ along the lines of~\cite{efetovbook} obtains 
 \begin{align*}
 {\cal C}_1(q,\eta)= & {8i\xi\over \eta t_H} {\rm Re}
 \int\limits_0^\infty dx \int\limits_0^x dy \, 
x  K_1(x)  K_{\sigma_q}(x) 
y K_1(y)   I_{\sigma_q}(y),
  \end{align*}
   where $\sigma_q={\sqrt{1-4iq\xi}}$, and $K_\nu$ and $I_\nu$ are the
 modified Bessel functions of order $\nu$. The $\sim
 \eta^{-1}$-scaling of this result implies a trivial (constant) time
 dependence $C_1(q,t)\equiv \mathcal{C}_\infty(q)$ at large times, where the
 saturation function, $\mathcal{C}_\infty(q)$,  is determined by the
 coordinate integrals.  The  isotropic component, $\mathcal{C}_0$, turns out to be
 given by the same expression, at, however, $q=0$. In other words, the forward
 scattering amplitude and
 the isotropic component coincide at large times,
 $\mathcal{C}_\mathrm{fs}(t=\infty)=\mathcal{C}_0(t=\infty)\equiv
 \mathcal{C}_\infty$, which means that the forward scattering peak  asymptotes
 to a value twice as large as the isotropic background. The saturation value
 $C_\infty(q)$ as a function of longitudinal momentum difference is shown in
 the left inset of Fig.~\ref{fig2}. The amplitude rapidly decreases as a
 function of $q$, half of the peak value $C_\infty$ is reached at the
 characteristic scale $q\simeq \xi^{-1}$, and for larger values the peak amplitude 
 decays as $\sim q^{-2}$.
 Since the localization length is much larger than the mean-free path $l$, 
 this narrow peak should be easily distinguished from the broadened energy 
 shell in $k$-space.

So far, we have assumed an initial state 
of sharply defined momentum. To 
account for the presence of momentum
 spread we convolute our results over a distribution of initial momenta. For
 a Lorentzian distribution  of
 width $\Delta k$, we find $\left\langle
 \mathcal{C}_\mathrm{fs}(\infty)\right\rangle = \mathcal{C}_1(i\Delta
 k,\infty)=\mathcal{C}_\infty(i\Delta k)$, where $\mathcal{C}_\infty$ is the
 peak function discussed above. The forward
 signal,  shown in
 the left inset of Fig.~\ref{fig2}, will be severely suppressed once
 $\Delta k >\xi^{-1}$. Qualitatively similar
 behavior is expected for other distributions, which means that 
 near monochromatic initial configurations are vital for the observability of
 the forward peak.

{\it Discussion:---}Previous experiments~\cite{Josse,Labeyrie2012} have probed 
the coherent response of atomic clouds to a speckle potential quench to 
two-dimensional disorder. In these systems, a crossover to effectively 
\emph{three}-dimensional dynamics 
(with only very weakly developed signatures of
quantum interference) occurred at rather short times $t\ll t_H$, which means that 
a coherent backscattering peak, but no forward peak could be observed.  
With this paper, we  propose  to repeat the
quench experiment in a quasi one-dimensional setting with its parametrically
shorter Heisenberg time, and stronger developed localization which will cause
a more rapid increase of the forward signal.
The quasi one-dimensional 
geometry is realized if $L_{y,z}\ll\xi\ll L_x$, where  the localization
length $\xi$ is of the order $N l$, and $N$ is the number of channels introduced 
by transverse size quantization.

For this type of system, our theory predicts the power law increase of a forward signal 
at short times, the saturation dynamics at large times, the momentum dependence of the 
forward signal, and its dependence on the width of the initial state. In total, this is
the first space/time resolved portrait of a strong localization phenomenon, with the perspective 
of observation using current experimental technology of cold atom physics or photonics~\cite{Segev}.

\acknowledgements T. M. would like to thank M. Micklitz for fruitful discussions. 
Work supported by FAPERJ (Tem\'atico and Infra 2013) and SFB/TR 12 of the 
Deutsche Forschungsgemeinschaft.



\begin{thebibliography}{99}
\bibitem{WeakField} 
$\mathrm{T}$ is already broken by fields of classically vanishing strength with a  
few flux quanta 
through the area $\mathcal{O}(L_\perp\xi)$, where 
$\xi$ is the localization length and $L_\perp$ the transverse width.
\bibitem{details} Further details will be published elsewhere.
\end{thebibliography}
\end{document}